\documentclass[conference]{IEEEtran}

\ifCLASSINFOpdf
  \usepackage[pdftex]{graphicx}
\else
\usepackage[dvips]{graphicx}
\usepackage{multirow}

\fi
\usepackage[english]{babel}
\usepackage{pslatex}

\usepackage{cite}
\usepackage{multibib}

\newcites{surv}{References}
\newcites{compl}{Published papers}

\begin{document}
\nocite{*}
\nocitesurv{*} 
\nocitecompl{*}

%
\title{Performance of Network and Service Monitoring Frameworks}

\author{\IEEEauthorblockN{Abdelkader Lahmadi and Laurent Andrey and Olivier Festor}
\IEEEauthorblockA{INRIA Nancy - Grand Est Research Center, Villers-L\`es-Nancy, France\\
Email: \{Abdelkader.Lahmadi,Laurent.Andrey,Olivier.Festor\}@loria.fr}
}
\maketitle
\begin{abstract}
 The efficiency and the performance of management systems is becoming 
  a hot research topic within the networks and services management community.
  This concern is due to the new challenges of large scale managed systems, where the
  management plane is integrated within the functional plane 
  and where management activities have to carry accurate and up-to-date information.
  
  We defined a set of primary and secondary metrics to measure
  the performance of a management approach. Secondary metrics are derived from the primary
  ones and quantifies mainly the efficiency, the scalability and the impact of management
  activities. To validate our proposals, we have designed and developed a benchmarking platform 
  dedicated to the measurement of the performance of a JMX manager-agent based
  management system. 

 The second part of our work deals with the collection of measurement data sets 
 from our JMX benchmarking platform. We mainly studied the effect of both load 
 and the number of agents on the scalability, the impact of management activities on the user perceived performance of a managed server and the delays of JMX operations when carrying variables values. 
Our findings show that most of these delays follow a Weibull statistical distribution. We used this statistical model to study the behavior of a monitoring  algorithm proposed in the literature, under heavy tail delays distribution.
 In this case, the view of the managed system on the manager side becomes
 noisy and out of date. 
\end{abstract}


\IEEEpeerreviewmaketitle

\section{Introduction}
The rapid growth of the Internet over the last 20 years
has been startling. However, efforts to manage its
services and their underlying networks have
often fallen afoul due to poor performance of the management 
systems in place. The problem is not that management systems and
protocols do not exist, but rather that the lack
of performance models, tools and benchmarking
platforms to assess their cost and well understand 
their needs on resource consumption are not well studied.
Furthermore, studying the performance of the value-added functional
plane without taking in consideration the cost of management activities
and its impact would lead to inaccurate estimation of the quality of service 
and might impact the business benefit.
Consequently, questions arise like: what is the cost of a management
system, its impact on a managed system ? how does a management
system scale with the growth of a managed system ?

The same problem has arisen in other computer
science disciplines (databases, distributed systems,
IP networks,etc). An extensive literature exists, 
and many standards have emerged in these disciplines to assess the performance
of the proposed systems and architectures. 

In the network and services management
community however, it was surprising that no
agreement on conventions for evaluating the performance of management systems
was established so far. Existing performance metrics like response times, throughput, cost, quality
and scalability in the network and service management literature are inconsistent and confusing.
As a result, no common foundation has been established
to evaluate the performance of management systems so far. 
Such studies are burdened with the lack of common metrics and usually their results are not
\textit{comparable}, their experiments are not \textit{reproducible} and they are not
\textit{representative}.

One approach
to solving the above lacks is to develop benchmarking platforms and collect
measurement data sets to identify and well define the most relevant
performance metrics and their measurement methodology.

The aim of our work was to provide common performance metrics to evaluate
the performance and the cost of management frameworks using both
measurement and analytical techniques for common unrealistic
and realistic management scenarios. We have focused our performance evaluation studies 
on JMX \cite{jsr003}, the \textit{de facto} standard to manage Java based applications.

The remainder of this paper is organised as follows. In section \ref{related}, we describe
related works devoted to the performance evaluation in general and those dedicated to the management 
plane in particular. In section \ref{statement},
we identify the major problems that the performance evaluation of
management frameworks studies did encounter. We detail our approach and our proposed set of metrics 
to evaluate the performance of a management framework in section \ref{approach}.   
In section \ref{results}, we present the main results that we have obtained. 
We conclude and address future work
in section \ref{conclusion}. 

\section{Related works}
\label{related}
Many studies have addressed the performance of management applications.
We classify these studies according the research goal behind them.
Mainly, we have identified three goals: comparing management approaches performance,
studying the performance of a new feature within an existing management approach,
purely studing the performance of a new management approach.
In \cite{pavlou:cm04,pras:etnsm04}, researchers   compare    SNMP
protocol  usage  based on polling  to  web service  based management.
In  an  effort  to   investigate the security  overhead   in  network
management, \cite{du:im01,corrente:noms04} assess the performance
of SNMP with security support. 
One of the rarely attempts to make a  dedicated measurement on a management protocol
was  undertaken   by  Pattinson \cite{pattinson:dsom01} for the SNMP protocol.
 
Performance evaluation is well studied  in other disciplines
for other purposes. 
In \cite{JAIN}, the author gives a cohesive collection on performance evaluation that include -- measurement techniques and tools, experimental design and analysis, simulation, queueing models and a section on probability theory and statistics. Our work is conducted in accordance to guidelines provided in this textbook.
The IPPM working group within the IETF has published many RFCs about performance 
metrics and their measurement methodology dedicated to IP networks \cite{ippm}. 
Their work inspired us to define and develop a set of metrics dedicated to monitoring applications.
The work of \textit{Woodside et al} \cite{woodside} has provided a metric to assess the scalability of 
a distributed system. We adopted this metric and extended it to assess the scalability and the impact of a monitoring application.
\section{Problem Statement}
\label{statement}
Our analysis of existing studies on the performance evaluation of management
frameworks, allows us to identify two major problems within these studies. 
The first problem is that \textbf{existing metrics are not adequate}.
Major metrics used by performance analysis studies have a confusing semantics. 
For example, monitoring delay is called  \textit{response time}, \textit{execution time}, \textit{round trip delay}, \textit{delivery delay}. Their metrics support only one activity and can not express others.
For example, a request-response delay metric cannot express notification delays.
These metrics, rather than giving a clear quantification of the performance, leave
a misunderstood burden on the performance of a monitoring application. The quantification of the scalability
is a good example.  
The second major problem in the current performance analysis is that \textbf{existing methodologies are not adequate}.
This problem engenderd the following flaws in existing performance studies:

\begin{itemize}

 \item    Results of  different      performance studies may  not   be
 \emph{comparable}.  For example, one study    may have measured   the
 delay of retrieving a  certain MIB table using \texttt{GetNext} PDUs,
 whereas the other study uses \texttt{GetBulk} PDUs.

 \item Experiments may  not  be \emph{reproducible}, which  means that
 other researchers trying  to perform the  same experiment may  not be
 able to  achieve similar results.  This problem  may be caused by the
 fact that the experiments  have  not been described with   sufficient
 details, leaving  the researcher trying  to reproduce the experiments
 with too many  options. This problem may  also be caused, however, by
 some events interfering  with the experiment, which
 basically means that the results are not correct.

 \item Performance studies   may not be \emph{representative},   which
 means that monitoring usage in real networks may  be largely different from
 the one assumed in a study.
\end{itemize}    
Therefore, we find that these problems originate from the lack of common 
measurement scenarios and metrics to evaluate a monitoring approach.

\section{Our Approach}
\label{approach}
We have developed an approach \cite{lahmadi:thesis} for the performance evaluation of a monitoring 
application, that relies on three goals. These goals are the following:
\begin{itemize}
 \item A well defined set of of performance metrics with a clear terminology. They should cover management challenges and they should be independent from the underlying technology;
\item A well defined methodology for their measurement and to guarantee reproducibility. 
\item Design a set of experimental procedures that cover monitoring challenges. The scalability and the development of benchmarking platforms we here targetted.   
\end{itemize}
To achieve these goals, we developed a set of contributions that we detail 
in the following sections.   

\subsection{A Unified Performance Metrics}
\label{metrics}
\begin{figure}[htp]
\centering
\scalebox{0.7}{\includegraphics{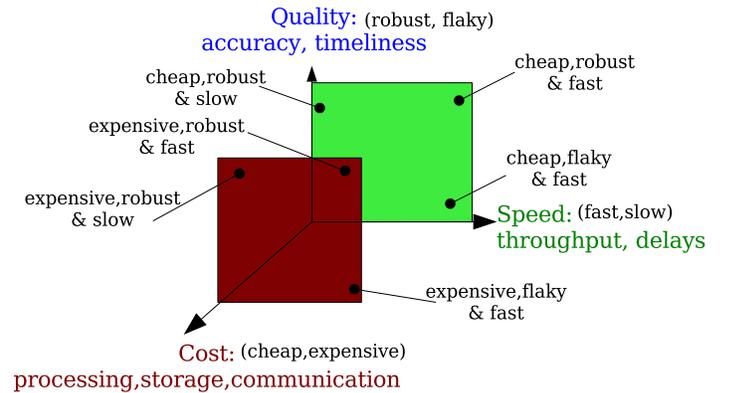}}
\caption{Definition space of primary performance metrics dedicated to monitoring frameworks evaluation.} 
\label{fig:metrics}
\end{figure}
We have defined tow categories of metrics to evaluate the performance of a monitoring application.

In the first category, we put a set of metrics that we qualify as \textit{primary metrics}.
These primary metrics are the direct output of a measurement process applied to a monitoring scenario under test.
As depicted in Figure \ref{fig:metrics}, this category includes three types of metrics which are: speed, cost and quality. 
The  {\it  speed} metrics  deal with management   data access  either
locally or remotely.  These  metrics capture the quantity and related
timing  of  management data  under some   management  activity.
The {\it cost} metrics deal with the overhead of management activities and the 
resources that they consume. The storage cost refers to  the amount of management data stored at 
a management entity. The communication cost refers to the total amount of information that
needs to be transmitted through the network between  two or many management
entities. The computational  cost reflects  the processing activity
at a management entity. The  {\it  quality} metrics  deal with  the  spatial and
 temporal  errors of management data.   By spatial error, we mean that
 we obtain an error    metric  per management attribute  that
 summarizes  its  deviation from  the  real  value of  the  manageable
 attribute.  The temporal error acts as an error metric for each time
 slot (a monitoring round, for example) summarizing the time deviation
 from that instant while acting  over a management attribute (collect,
 notify, update).   We   note, that   a fundamental  trade-off  exists
 between the three categories of metrics.
These primary metrics need to be extended when they are measured in a 
particular monitoring scenario involving one to one management entities 
or many management entities. Therefore, we define two types of qualifiers
to be used of these metrics according to the involved monitoring scenario.
These qualifiers are the following:
\begin{itemize}
\item One-to-one: collects the measured metric vector in a manager to an agent scenario;
\item One-to-many: collects the set of singleton measurement between a manager and many agents;
\end{itemize}
When the monitoring scenario involves a manager, an intermediate and an agent, 
the measured metrics needs to be decomposed to quantify the per-entity contribution
to the end-to-end performance.

The second category of metrics are qualified as \textit{derived metrics}.
These derived metrics are computed from the measured primary metrics.
The motivation behind the definition of these metrics is to quantify the efficiency of a monitoring 
approach within a single quantity in order to be able to compare them.
The efficiency metric of a monitoring application puts in relation the three types 
of primary metrics that we have defined above. Thus, we obtain the following formulae:
\begin{equation}
\label{eq:efficiency}
G(k)=\frac{R(k)}{C(k)}xQ(k) 
\end{equation}
 Where $G(k)$ is the efficiency quantity computed under a performance factor that takes a
value denoted by $k$. $R(k)$ denotes a speed metric quantity, $C(k)$ denotes a cost metric quantity
and $Q(k)$ denotes a quality metric quantity.
For example, to compute this efforts we can use the throughput of a monitoring system in terms of the number of collected attributes per second as a speed metric, the cost of management activities in terms
of resource consumption (CPU, memory and network bandwidth) as a cost metric 
and the quality of monitoring operations in terms of their respect of a tolerable
delay.

The definition of the efficiency derived metric will allow us to define more 
metrics that cover monitoring challenges which are the scalability and the impact of
a monitoring application. The scalability metric will be detailed in section 
\ref{scalability} and the impact of a monitoring  activity on the performance
of the managed system will be detailed in section \ref{impact}. 

\subsection{A Benchmarking platform}
To better understand the performance of management frameworks, we have
developed a benchmarking platform \cite{RT-laurent} 
dedicated to JMX management paradigms. We selected this framework 
because it provides inherent manageability to Java technology enabled 
applications and services, therefore, the impact of management activities is
more visible. The JMX framework has not a predefined management information model,
instead it offers several types of managed objects (MBeans) 
to instrument resources and supports many protocols (TCP/RMI, HTTP, SOAP) 
for the communication between a manager and an agent.
Nevertheless, our platform is flexible and modular enough to be 
extended to other management frameworks like SNMP.
We have focused on the manager-agent model with synthetic tests where we have varied the number of managed objects (MBeans), their types, the monitoring rates and the number of agents. 
Despite the high saleability of benchmarking results
\cite{JAIN} which is its key justification, our experience on
the benchmarking of the JMX framework, shows us how much this technique
is time consuming and error-prone. It also reveals its limited coverage
of the space of performance factors values. For example, in \cite{lahmadi:aims07},
a complete coverage of all our measurement series to identify 
the impact of the three integration models of an agent within a managed
systems (as depicted in Figure \ref{fig:driver-impact}) needed 3 months of measurement, or 1 month at best if we parallelize measurements.
Thus, we believe that analytical and simulation techniques are more
suitable to investigate deeper the performance of management frameworks.
However, before doing simulation we need to model the behaviour of
management activities.
\begin{figure*}[htp]
\centering
\scalebox{0.4}{\includegraphics{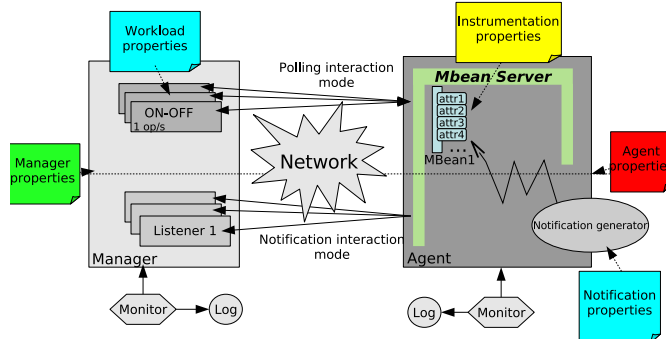}}
\caption{Architecture of the JMX-based instance of our benchmarking platform.} 
\label{fig:platform.benchmarking}
\end{figure*}


\section{Results}
\label{results}
In this section, we present the main results that we have obtained by using our
benchmarking platform and the set of performance metrics that we have defined.
\subsection{Monitoring Impact}
\label{impact}
We have built a derived metric to compute the impact of a monitoring activity
on the performance of the managed system. This metric has implications on network management 
models like the SNMP, as well as on services and applications management frameworks.

To develop this metric, we used the efficiency metric
described by equation \ref{eq:efficiency}. We denote by $E(k)$ the productivity of
a managed system. The productivity is defined as:
\begin{equation}
E(k)=\frac{F(k)}{F(k)+G(k)} 
\end{equation}
Where $F(k)$ is the managed system efficiency defined in the same manner as the monitoring efficiency
$G(k)$ and computed by equation \ref{eq:efficiency}. 
This productivity captures the ratio between the functional efficiency of a managed system defined by 
the real work that offers to the users and the real efficiency that it needs to accomplish this work.
Therefore, we define the Management Impact Metric as follows.
\begin{equation}
MIM(k_0,k)=1 - \frac{E(k)}{E(k_0} \in [0,1] 
\end{equation}
 Where $k$ denotes an impact factor that can be the number of monitoring requests.
In \cite{lahmadi:dsom05}, we illustrated this metric on the monitoring of 
a managed J2EE server\footnote{We used JBoss as managed server: http://www.jboss.org} JBoss using the JMX framework. In this case, we varied the monitoring rate as an impact factor. 

\begin{figure}[htp]
\centering
$\begin{array}{cc}
\scalebox{0.33}{\includegraphics{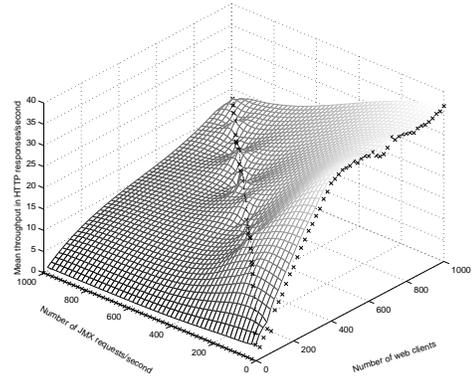}} \\
\mbox{\bf (a) } \\
\scalebox{0.33}{\includegraphics{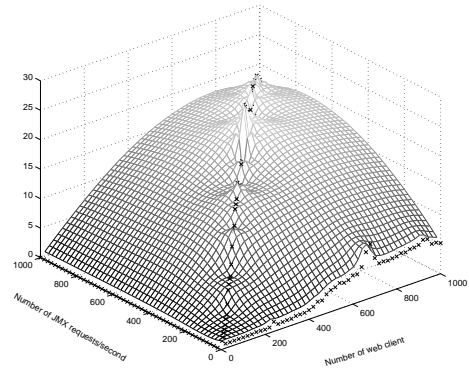}} \\
\mbox{\bf (b) }
\end{array}$
\begin{center}
\caption{The impact of a driver model of a monitoring agent 
on a web server performance in terms of (a) throughput and (b) 
HTTP transactions delays.}
\label{fig:driver-impact}
\end{center}
\end{figure}

In a second stage, we have extended this initial work to study the impact of instrumentation models as described in \cite{Kreger} on both the performance of the management and an instrumented web server
\cite{lahmadi:aims07}.
We did show that the users perceived performance
in terms of the number of HTTP transactions/s and their respective delays are highly affected by the management activity in the boot driver and component models while a daemon integration model limits the management activities impact on the functional plane. However, we showed that under low monitoring rates in the order of 200 requests/second, the three integration models have a small impact on the web server performance. 
Figure \ref{fig:driver-impact} depicts the throughput and the delays of the web server, monitored 
by a driver agent model. In this experiment, we varied the operational load in terms of number of web clients   
and the monitoring load in terms of \textit{getAttributes} per second. 

\subsection{Monitoring Scalability Assessment}
\label{scalability}
\begin{figure}[htp]
\centering
\scalebox{0.4}{\includegraphics{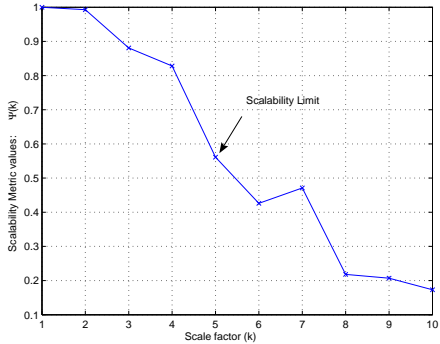}}
\caption{Scalability assessment of the manager-many agents model, where we varied the number of agents as a scale factor between 70 to 700.} 
\label{fig:scalability}
\end{figure}

The development of a monitoring efficiency metric that we described in section 
\ref{metrics} allowed us to the define a unique metric to assess the scalability
of a monitoring application. This metric relies on the one proposed in \cite{woodside}
to quantify the scalability of distributed systems. 
Therefore, we define the scalability degree $\Psi(k_1,k_2)$ of a monitoring
as follows.
\begin{equation}
\Psi(k_1,k_2)=\frac{G(k_2)}{G_k1}
\end{equation}
where $k_1$ and $k_2$ are two different values of a scale factor.
Generally, $k_1$ is fixed to an appropriate value to compute a baseline efficiency
of a monitoring application. The value of $k_2$ will vary and the scalability degree is computed to capture 
the degradation or enhancement of the monitoring application scalability.
A monitoring application scales well when $\Psi(k_1,k_2) \geq 1$ or 
$\Psi(k_1,k_2) \approx 1$.
In \cite{lahmadi:cfip06}, we have illustrated this metric on the scalability
assessment of a centralized monitoring application involving one manager and 
many agents. We studied its scalability degree analytically by using the performance model
developed in \cite{chen:jsac02}. We found analytically that the scalability
limit is mainly influenced by the monitoring delays. Thus, if the delays remain
constant, the scalability degree stays close to 1. We also found that network transit delays
when they become higher, contribute to enhance the scalability of such an approach. When the number of agents becomes high, the network will delay the responses from the agents 
and decrease the load on manager. However, this network transit delay has to be bound
to keep an acceptable monitoring quality, mainly the timeliness.
Figure \ref{fig:scalability} depicts the scalability degree of a JMX-based monitoring
application while we varied the number of agents between 70 and 700 agents by a step of 70.
We observe that the scalability limit of the monitoring application is close to 350 agents
that corresponds to a scale factor of 5.
   
\subsection{Monitoring Delays Characterisation}
When assessing the impact and the scalability of a monitoring application,
the primary key metric is the monitoring delays.
We define the  monitoring delay as the delay that experiences a monitoring attribute 
within an activity, to retrieve/alter its value.
This delay is measured using the one-to-one, one-to-many metrics defined
in section \ref{metrics}. 
Based on the developed benchmarking platform, we have elaborated in \cite{lahmadi:dsom06},
a delay model for the manager-agent paradigm. We have collected measurements within two main scenarios.
The first scenario use a single manager/single agent setup, where we used the one-to-one delay metric. 
The second scenario uses a single manager/multiple agents setup, where we used the one-to-many metric. 
For both scenarios, we applied statistical analysis on the collected data sets to identify their 
underlying statistical distributions.
In the first scenario, we varied the monitoring rate between 1 and 1000 polled attributes per second.
We find that the one-to-one delay closely follows a \textit{LogNormal} distribution. 
In the second scenario, we varied the number of agents between 70 and 700 agents.
We found that the one-to-many delays closely follow a normal distribution with a small number of agents, 
and becomes more heavy tailed and approximates a \textit{Weibull} distribution 
with a considerable number of agents. 

\begin{figure}[htp]
\centering
\scalebox{0.4}{\includegraphics{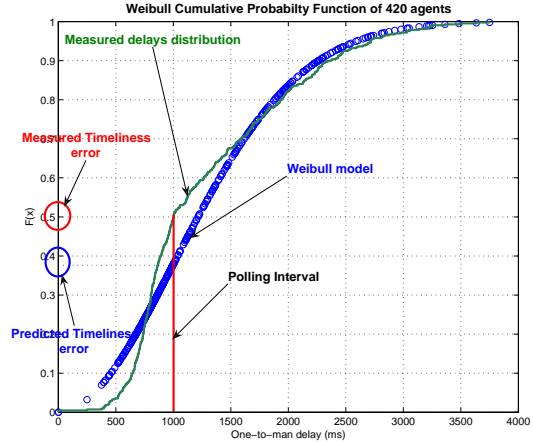}}
\caption{Predicted and measured timeliness of a polling activity with one manager and 420 agents.} 
\label{fig:timeliness}
\end{figure}

The delays scaling behaviour
is interesting because we can quantify the temporal accuracy 
of management data collected by a manager to a somewhat
{\it delay tolerance}. This parameter is considered as an upper bound
on the delay. Any collected attribute from an agent
experiences delays, that hold with high probability and is determined empirically based on the managed environment. 
Figure \ref{fig:timeliness} depicts the measured and predicted timeliness of 
a polling activity involving one manager and 420 agents. The timeliness is defined as the fraction
of agents that respond with delays lower than the monitoring interval which is equal to 1 second in this case.
We observe that the measured and predicted timeliness are close to 0.5 which means that half of the agents
respond with delays greater than 1 second.  
The methodology uses to model management delays can be applied to other management related performance metrics 
such as the number of polled/notified attributes.

\begin{figure*}[htp]
\centering
\scalebox{0.4}{\includegraphics{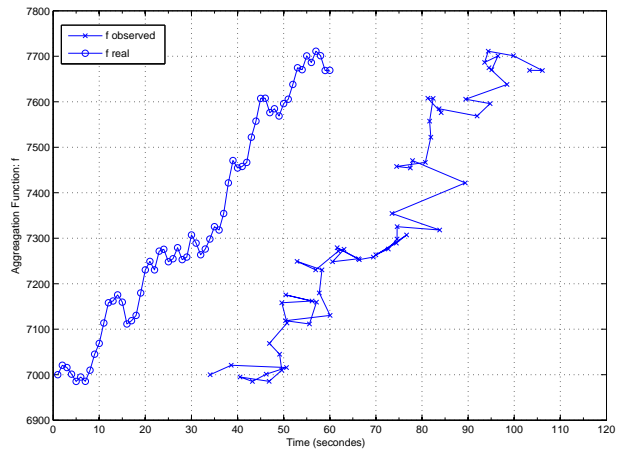}}
\caption{Effect of delays on the behavior of a monitoring algorithm with 700 agents attached to the manager in 
a polling mode interaction.} 
\label{fig:distorsion}
\end{figure*}

After the identification of the statistical distribution of monitoring delays
within JMX-based applications, we used the empirically obtained models to simulate the effect
of these delays on the behavior of the monitoring algorithm running on the manager node.
We developed Matlab scripts to simulate a centralized monitoring algorithm behavior
by generating synthetic delays from a Weibull distribution. The parameters of the 
distribution are obtained from our measurements as described above.
We used a monitoring algorithm that computes an aggregation function using single values
from many agents. This algorithm is similar to the simple-rate algorithm described in \cite{raz:infocom01}.
For each simulation, we fix the number of agents, we generate the synthetic delays and we
compute the real aggregation function from the values remaining on the agents and the observed one
computed on the manager. We find that the discrepancy between the observed aggregation
function computed on the manager and the real one computed without monitoring delays 
is very important when the number of agents is close to 700.
Thus, the monitoring view on the manager experiences a temporal distortion when the number
of agents increases. This is due to the monitoring delays.
Figure \ref{fig:distorsion} depicts the result of the simulation of the monitoring algorithm
with a number of agents equal to 700 and the synthetic delays generated from a Weibull distribution.

\section{Conclusion and future works}
\label{conclusion}
The aim of our work is to provide a common evaluation methodology and 
performance assessment metrics of a management framework. 
For this purpose, we defined a set of primary metrics to capture the speed, the cost and the quality
of management operations. We designed and developed a benchmarking platform
to measure these metrics within realistic and unrealistic scenarios.
We illustrated our work with the manager-agent JMX-based management framework, widely used to manage
$J2EE$ applications on which several Internet services rely (retailers, banks, government institutes,etc). 
Our findings show that the management activities have a deep impact
on the performance of a managed service if they are optimized. 
This impact depends on the monitoring rate 
and the integration model of the agent within the managed system. 
We also show that high monitoring rates degrade both the performance
of management and managed systems, especially delays that become more random
and their underlying statistical distribution more heavy-tailed. 
Thus, we believe that optimizing management activities by minimizing their cost and rates while maximizing their coverage and business benefit needs to be fitted and well defined  within solid optimization frameworks.

\bibliographystylesurv{IEEEtran}
\bibliographysurv{surveyed}

\bibliographystylecompl{IEEEtran}
\bibliographycompl{compl}

\end{document}